\documentclass[wcp]{jmlr}

\usepackage{listings}
\newcommand{\code}[1]{{\small\textsf{#1}}}

\usepackage{longtable}
\usepackage{booktabs}

\usepackage{lineno}
\makeatletter
\let\Ginclude@graphics\@org@Ginclude@graphics 
\makeatother

\jmlrvolume{1}
\jmlryear{2024}
\jmlrworkshop{2024}

\begin{document}

\title{Defect Prediction with Content-based Features}

\author{
\Name{Hung Viet Pham} \Email{hvpham@yorku.ca}\\
\addr York University, Canada
\AND
\Name{Tung Thanh Nguyen} \Email{tung@tamu.edu}\\
\addr Texas A\&M University, USA
}

\maketitle
\begin{abstract}
Traditional defect prediction approaches often use metrics that measure the complexity of the design or implementing code of a software system, such as the number of lines of code in a source file. In this paper, we explore a different approach based on content of source code. Our key assumption is that source code of a software system contains information about its technical aspects and those aspects might have different levels of defect-proneness. Thus, content-based features such as words, topics, data types, and package names extracted from a source code file could be used to predict its defects. We have performed an extensive empirical evaluation and found that: i) such content-based features have higher predictive power than code complexity metrics and ii) the use of feature selection, reduction, and combination further improves the prediction performance.
\end{abstract}

\keywords{Defect prediction, text analysis, code analysis}

\section{Introduction}
Software defects occur frequently in software development and often lead to costly and time-consuming activities to find and fix them. \cite{nist2002} report that software defects cost the US economy nearly \$60 billions a year. In addition, \cite{ibm2002} found that finding and fixing them accounts for 50 - 75\% of the total development cost in a software project. The later a defect is detected in the life cycle of a software product, the higher cost and effort are needed to fix it and re-deploy the fixed version of that product back to the field.

Many methods, techniques, and tools have been developed to support the early detection of software defects. One important line of research is {\em defect prediction}. Defect prediction approaches aim to identify the most defect-prone modules (binaries, source files, classes, or functions) in a given software system. Such prediction results can help software engineers to focus their manual defect detection effort like code review or testing on modules with higher likelihood of success, thus, improving the effectiveness and reducing the cost of their activities. Extensive literature reviews on existing defect prediction approaches can be found in~\cite{hall2011,shihab}.

Most researches on defect prediction focus on the factors and metrics could be used to predict defects. Often called \emph{predictors} or \emph{features}, they are used as input of a \emph{prediction model} which outputs the predicted number of undiscovered defects in a software module (regression models) or whether it is defective (classification models). These factors are general in nature and follow a common belief about software systems.

Traditional defect prediction approaches for source code often use metrics measuring the complexity of the design or the implementing code of a software system. For example, the most commonly used code metric is the number of lines of code (LOC). Other frequently used object-oriented design metrics are the depth of inheritance tree (DIT), the number of children (NOC), the lack of cohesion in methods (LCOM), or the coupling between objects (CBO) of a class.

In this paper, we explore a new approach based on two key assumptions. The first one is that a software system often implements several groups of functionality, each might have different levels of defect-proneness. For example, JEdit is a subject system studied in this paper. It is a word processor (document editor) with functionality for managing graphical user interface (GUI), presenting documents, processing edit commands from users, searching text, managing files (e.g. loading, saving, parsing), etc. Our study of JEdit suggests that while the code for GUI and editing commands is highly defect-prone (e.g. the most defective file is \code{JEditTextArea.java} with up to 45 post-release defects), the code for text search and parsing is much less defect-prone. Therefore, if we can infer the functionality implemented in a code module (e.g. a source file or a class) and the defect-proneness levels of such functionality from historical data, we can predict defects in that module.

We make the second assumption in our approach that functionality implemented in a code module could be inferred from its content, i.e. from identifiers, comments, annotations, string literals, keywords, embedded documentation, etc. For example, developers often name classes and methods using identifiers suggesting their functionality. For example, JEdit has some classes named \code{JEditTextArea}, \code{OptionsDialog}, and \code{Buffer} which clearly indicate the functionality they implement. In code comments, developers can also explain and discuss the functionality of their code such as the implemented algorithms or the roles of variables and parameters.

Based on those ideas, we explore four new types of features extracted from code content for defect prediction. The first one includes textual \textbf{terms} extracted from all text elements in code like comments or identifiers. We use standard tokenizing and stemming techniques to extract those terms and use the bag-of-word model to represent them. For example, the identifier \code{OptionsDialog} is tokenzied into two terms \code{Options} and \code{Dialog} which are further stemmed as \code{opt} and \code{dialog}, respectively.

Although textual content of source code could contain most information of the implemented functionality, the amount of extracted terms can be large and noisy. Therefore, we use topic modeling~\cite{blei03} as a feature reduction technique for the extracted terms. This produces \textbf{topics}, the second type of features explored in this paper. Prior studies suggests that topics extracted from source code of a software system correspond to its technical concerns (e.g. functionality) and those topics can be used to predict defects~\cite{nier11}. For example, in JDT, a compiler framework, code written for \emph{semantic analysis} is more likely to have errors than code written for \emph{lexical analysis}.

To further address the noise in \emph{textual features} like terms and topics, we investigate \emph{programming-semantic features} including the data \textbf{types} of variables and objects, and the \textbf{packages} containing those types. Compared to text features, types provide a higher level of abstraction. In object-oriented design, a particular data type, e.g. a class, is defined to perform certain tasks, which contributes to the actual functionality of the code using that type. For example, objects of type \code{java.io.File} represent particular files or folders. They provide methods to obtain information about files and to manipulate those files in the system. Hence, if a source file uses \code{java.io.File} type, it likely works with the file system (e.g. create, delete, read, or write files) or performs file I/O functionality. In contrast, objects of type \code{org.eclipse.jdt.core.dom.ASTParser} provide functions to parse Java code in abstract syntax trees. Thus, if a source file contains objects of this type, it is likely to have parsing/compiling functionality. That means, we could infer functionality of a code unit based on the presence of some particular data types.

As a large software system might contain thousands data types, we consider package organization as a feature reduction technique for type features. In object-oriented programming, package organization can improve code modularity by grouping related modules (e.g. classes, source files) into packages. As related classes are grouped together, the resulting packages can provide higher levels of abstraction. For example, in Java API libraries, package \code{java.io} provides input/output functions while package \code{javax.sql} provides data access and processing functions. Like modeling infers the technical concerns of source code from textual features, package organization could infer technical concerns from data type information. However, while topic modeling is an unsupervised learning task, package organization was done by human, i.e. developers that design the target system, hence it would be more accurate.

While extracting text, topic, type and package features, we construct the feature vectors for a source file using the count of each term, the (log-transformed) counts of terms assigned to each topic, the presence of each type, and the presence of each package, respectively. Those vectors can further be combined into a unified one for all types of features.

As a software system can has high numbers of the four newly proposed features, we apply feature selection techniques to improve the prediction performance, in both accuracy and running time. For example, we only select terms and type features that have high ranked correlation or mutual information with defects. We also apply principal component analysis (PCA), a standard feature reduction technique to reduce the potential correlations among selected features.

Because this paper focuses on the features rather than the models for defect prediction, we only use the simple linear regression model (LR) for the prediction task. It should be noted that other prediction models like logistic regression, decision tree, or support vector machine, can also be used.

% RESULT

We have conducted an extensive empirical evaluation on a public defect dataset including 42 releases of 14 real-world software systems. This evaluation contains more than 2,000 experiment runs, exercising different options of the prediction model, such as the number of selected term features, the number of topics extracted, the number of selected type features, etc. We also compared our features with the best reported traditional code metrics.

The results show that all our proposed features are predictive of defect-proneness and provides better prediction results than traditional code metrics like number of lines of code (LOC) or Chidamber-Kemerer (CK) metrics. More importantly, feature selection and reduction techniques could further improve prediction accuracy. Finally, combining all four types of features is better than using them alone.

The key contributions of this paper include:

1. New types of defect predictors including terms, topics, types, and packages extracted from source code, and

2. An empirical evaluation to compare those predictors with traditional code metrics.

In Section 2, we introduce our features in details, including techniques for extracting them from source code and selecting the best ones. Section 3 describes our evaluation settings and Section 4 reports its results. 
%In Section 5, we discuss a detailed analysis of our proposed features in two representative systems and our plan for future explorations. 
Section 5 presents related work and conclusions appear last.

\section{Approach}

In this section, we describe in details the extraction process for term, topic, type, and package features. We also discuss several feature selection and reduction techniques for those features.
While term features can be extracted using simple text analysis techniques, topic features are inferred using LDA, a widely used topic modeling technique by \cite{blei03}. Extracting types and packages is more complicated, involving code parsing and partial program analysis (PPA).

\subsection{Extracting term features}

Term features are extracted directly from the textual content of source code. First, raw code was tokenized using whitespace, numeric, and special characters as separators. Because our subject systems are all Java projects, each resulted token is further split using the Java name convention (i.e. camel casing). For example, token \code{StringBuffer} is split into two words \code{String} and \code{Buffer}; token \code{JEditArea} is split into three words, \code{J}, \code{Edit}, and \code{Area}. After that, words of length 1 like \code{i} or \code{J} are disregarded. The remaining are lowercased and stemmed using the standard Porter stemmer. For example, \code{Buffer} becomes \code{buff} and \code{Condition} becomes \code{condit}. Finally, the vector representing term features for each source file is constructed using bag-of-word model and weighted using \emph{tf.idf} scheme.

%Across all test projects, listed in Figure~\ref{tab:system}, a typical term features vector has a size of around 2500. This number can reach up over 11000, in Xalan 2.7 and 2.6, while it can be as low as 1100, in Synapse 1.0. Even with only 1100 features, the feature vectors are still quite large and would include features that will only introducing noise but without any added benefits. To solve this problem, we applied several feature selection and input space reduction techniques that bases on Spearman and Pearson correlation coefficient (SCC and PCC), Mutual Information (MI), and Principal Component Analysis (PCA). These techniques will be described more in details later in section \nameref{sec:featureSelection}.
	
\subsection{Extracting topic features}

In our work, topic features are extracted using LDA~\cite{blei03}. By applying LDA, our approach assumes each software system to have $D$ source files, $V$ words, and $K$ topics. Each topic is a distribution over all those words and is a sample of the Dirichlet distribution $\rm{Dir}(\beta, V)$. Each source file has a distinct topic proportion which is a sample of Dirichlet distribution $\rm{Dir}(\alpha, K)$. Each word in a source file is assigned to a topic.

For example, assume that JEdit, an editor, has only two topics \emph{``text editing''} (ED) and \emph{``graphical user interface''} (GUI). LDA assumes each word has a probability to be assigned to each of those two topics. However, the probabilities of assigning words \code{edit},  \code{delete}, or \code{buffer} to ED are higher than to GUI. In contrast, the probabilities of assigning words \code{button}, \code{window}, or \code{dialog} to GUI are higher. Source file \code{JEditArea.java} could have 70\% of its words assigned to ED and 30\% to GUI. A word \code{view} in this file is likely assigned to GUI.

LDA is an unsupervised technique which automatically infers topics from documents. Its input includes $D$ documents, $V$ words, and the number of topics $K$. Its output is $K$ distributions $\phi_{k=1..K}$, each for a topic and $D$ topic assignment vectors $\varphi_{d=1..D}$. That is, $\phi_k(w)$ is the probability word $w$ is assigned to topic $k$, while $\varphi_d(k)$ is the number of words in document $d$ assigned to topic $k$.

After applying LDA on the source files of a system, we log-transform the topic assignment vector of each file to construct its topic feature vector, expecting that log transformation will reduce the unbalance between common words and rare words. One could consider topic modeling as a reduction technique for term features, as we can produce topic feature vectors of $K$ dimensions from term feature vectors of $V$ dimensions.
	
\subsection{Extracting type features}

To extract type features in a source file, we first parse it into an abstract syntax tree. Then, we resolve type-binding for any identifier, object, variable, and expression appearing in the tree. To make the type-binding resolution robust, we employ Partial Program Analysis, which can perform type-binding even when the system is not completely compiled or has some missing dependencies.
 
Rather than counting the occurrences like for term features, the feature vector for type features is binary, i.e. it denotes only the presence/appearance of a type in a source file. This design decision is suggested by our preliminary investigation, when binary vectors for type features outperform the corresponding count vectors in defect prediction. 

\subsection{Extracting package features}

The type binding process provides qualified name for all resolved types. For example, a variable \code{parser} is resolved to have type \code{org.eclipse.jdt.core.dom.ASTParser}. This qualified name indicates this type belong to the package \code{org.eclipse.jdt.core.dom}. 

Thus, this package is considered to appear in the source file containing that variable. This suggests us to extract package features from type features for each source file.

Similar to type feature vectors, we also construct package feature vectors in binary representation because our preliminary investigation suggests that it proves better prediction results. That means, the package feature vector of a source file only indicates if a type of a package is used in that source file or not. One could consider package features as a reduction of type features, as a package often contain several types.

\subsection{Feature selection and reduction}
\label{sec:featureSelection}

The feature spaces for term and type features, are generally large. For example, Xalan 2.7 has 11,087 extracted term features and MYL has 4,004 extracted type features. It is likely that many of those features are noise. For example, common words like \code{an} or \code{the} or common types like \code{int} or \code{String} appear frequently in source code and do not have high correlation to defects. In addition, related words or types often go together, thus the corresponding terms can be highly correlated. Thus, to reduce those noisy and highly correlated features, we use several feature selection and reduction techniques.

A feature selection method works by i) computing a relevance score between each feature and the defect count in the data, ii) using those scores to rank the corresponding features, and iii) selecting features with highest scores. Following prior studies in defect prediction and machine learning, we use three kinds of scores: Pearson correlation coefficients (Pearson), Spearman correlation coefficients (Spearman), and Mutual Information (MI). %Detailed formula for those scores could be referred in~\cite{jerome03}.

The selected features might still be correlated. We address that situation by using Principal Component Analysis (PCA), a standard feature reduction technique. PCA transforms a potential correlated feature set into a set of \emph{principal components} (PC) that are linearly orthogonal. By selecting top principal components that account for most (typically 90\%) variance in the data, we could remove overlapping information and reduce noise. 
\section{Empirical Evaluation}

\subsection{Datasets}

\begin{table}
\tiny\textsf
	\centering
	\caption{Subject Systems}
	\begin{tabular}{llrrrrr}
		\addlinespace
		\toprule
		ID    & Name  & Version & No. Files & No. Term & No. Type & No. Package\\
		
		\midrule
		ALR set& 4 projects & 4 releases &\\
		
		\midrule
		JDT   & Eclipse JDT Core & 3.4   & 995 & 6,120 & 3,052 & 45\\
		PDE   & Eclipse PDE UI & 3.4.1 & 975 & 4,052 & 3,527 & 66\\
		MYL   & Mylyn & 3.1   & \textbf{1,063} & 4,510 & \textbf{4,004} & \textbf{127}\\
		EQU   & Eclipse Equinox framework & 3.4   & 322 & 3,967 & 1,019 & 86\\
		%LUC   & Apache Lucene & 2.4.0 & 641 \\
		
		\midrule
		JM set& 11 projects & 38 releases &\\
		
		\midrule
		Ant	& Apache Ant & 1.3 - 1.7 & 123 - 740 & 2,040 - 4,926 & 346 - 1,558 & \textbf{8} - 67\\
		Camel & Apache Camel & 1.0 - 1.6 & 333 - 927 & 1,325 - 2,344 & 1,182 - 2,970 & 46 - 120 \\
		%Forrest & Apache Forrest & 0.7 - 0.8&\\
		Ivy & Apache Ivy & 2.0 & 352 & 1,986 & 743 & 52\\
		JEdit & JEdit & 3.2.1 - 4.2 & 260 - 355 & 2,744 - 3,371 & 882 - 1,241 & 16 - 23\\
		Log4J & Apache Log4J & 1.0 - 1.2 & \textbf{104} - 194 & 1,852 - 2,282 & 287 - 600 & 12 - 25\\
		Lucene & Apache Lucene & 2.0 - 2.4 & 186 - 330 & 1,936 - 2,650 & 378 - 655 & 10 - 13\\
		%PBeans & pBeans & 1.0 - 2.0&\\
		Poi & Apache POI & 1.5 - 3.0 & 234 - 437 & 2,442 - 3,913 & 459 - 821 & 19 - 20\\
		Synapse & Apache Synapse & 1.0 - 1.2 & 157 - 256 & \textbf{1,180} - 1,690 & 494 - 924 & 23 - 33\\
		%Tomcat & Apache Tomcat & 6.0.389418&\\
		Velocity & Apache Velocity & 1.4-1.61 & 195 - 229 & 2,279 - 2,335 & 424 - 476 & 25 - 30\\
		Xalan & Apache Xalan-Java & 2.4 - 2.7 & 676 - 899 & 4,716 - \textbf{11,087} & 1,245 - 1,565 & 38 - 42\\
		Xerces & Apache Xerces & init - 1.4.4 & 162 - 452 & 2,299 - 3471 & \textbf{270} - 706 & 17 - 28\\
		
		\bottomrule
	\end{tabular}%
	\label{tab:system}%
\end{table}%

To evaluate the predictive power of our proposed features in comparison to the traditional ones, we conducted several experiments on 42 releases of 15 open-source projects. Table~\ref{tab:system} provides a brief description of those systems, including their versions and the number of collected source files. The bug data was included in two different publicly available datasets provided in prior work. The first dataset (ALR set) includes bug data, CK and OO metrics extracted for four open-source software systems and the second (JM set) provides bug data and CK metrics for 38 releases of 11 projects. Lucene 2.4 is included in both datasets so to maintain consistency it was excluded from the first dataset. JM set also includes Apache Forrest, pBean,  Apache Tomcat, and JEdit 4.3 but these were removed due to their small size (less than 100 files) or their lack of reported bugs (the bug ratio less than 10\%). These datasets do not contain source code of those systems, we retrieved the code directly from their source repositories using the provided version information.

\subsection{Evaluation method}

We performed several experiments to evaluate the effectiveness of different feature types in predicting defects and provided a general estimation for parameters of each feature type. This section will describe our experiment methods and settings. Features were extracted using code written in Java and experiment steps including feature selection, model training, evaluation, and qualitative analysis were done using code written in Matlab and R.

\subsubsection{The prediction model}
To predict defects, we used a Linear Regression model. The reasons to select such a simple model are: i) we mainly focus on evaluating our proposed features, by selecting a simple universal and easy to train model we could quickly evaluate our system, ii) we want to compare our proposed features with other metrics that have been proposed of which many were implemented using this simple model. \cite{shihab} provided an extensive list of models used in defect prediction of which LR is the second most frequently used model after Logistic Regression. LR model works based on an assumption that the input features are not highly correlated so our PCA step makes sure that redundant information and noise was removed.

\subsubsection{Performance measurement and cross validation}
To measure the system performance, we used two metrics: Spearman ranked correlation coefficient (SCC) and Mean Absolute Error (MAE). SCC measures the ranked correlation coefficient between the predicted defects and the actual numbers of post release defects, the higher the coefficient the better the system at predicting defect-proneness of source code. Because we focus our effort in creating a system that could rank source file in term of its defect-proneness, SCC metric is an important performance measure that we consider. MAE metric is used to evaluate the system's ability to predict the actual number of defects in a source file by measuring the average difference between the predicted and actual number of defects. Thus, a lower MAE indicates the better prediction performance.
%mean absolute prediction error.

We used cross validation such that each experiment was repeated 50 times using 90\% data for training and 10\% for testing. To make paired t-tests valid when comparing different features and selection methods across all projects, a fixed random seed was set before each experiment so the same subject system will be cross-validated using the same fold configurations for each feature or method.

\subsubsection{Baseline}
We compared our result with existing metrics that has been used in the past:  the traditional and simple Lines of Code (LOC) and the more recent CK metrics. To train the baseline models, we used pre-compiled LOC and CK metrics provided in the datasets. The first four projects in the ARL set also contains some Object Oriented (OO) metrics. We include them in the evaluation of baseline systems for those projects as well.

\section{Experiments and results}

We conducted five experiments to evaluate our proposed features and methods of feature selections. The first fours examined each feature type in turn, the final experiment was conducted to evaluate combined features. Each of the first three experiments was split into phases to discover different optimal settings for each feature type. Every experiments were performed on all 42 released of 15 projects. To evaluate baseline systems, in the final experiment, we used the CK and OO metrics provided for some projects in the ALR set while for all others we used only available CK metrics.

\subsection{Experiment 1: Term features and term feature selections}

In this experiment, we evaluated our proposed term features' performances and discover the most optimal configuration for term features including which feature selection method and what number of selected term features should be used. In the first phase of this experiment we ran our system using all three methods of selection: Spearman, Pearson, and MI with the number of term features selected of 5, 10, and 20 to determine the most appropriate selection method. We also ran with no feature selection, i.e. all terms are selected.

\begin{table}
\small\textsf
	\centering
	\caption{Prediction performance of selection methods for term features}
	\begin{tabular}{ l rr rr rr rr }
		\addlinespace
		\toprule
		
		Selection method & \multicolumn{2}{c}{Spearman} & \multicolumn{2}{c}{Pearson} & \multicolumn{2}{c}{MI} & \multicolumn{2}{c}{All features} \\
		Prediction measure & SCC & MAE & SCC & MAE & SCC & MAE & SCC & MAE \\
		
		\midrule
		
		%\multicolumn{9}{l}{Average across all releases}\\
		
		%\midrule
		
		Across all releases         & \textbf{0.461}  & \textbf{0.676}  & 0.432  & 1.088  & 0.397  & \textbf{0.676}  & 0.352  & 1.510 \\
		
		\midrule
		
		\multicolumn{9}{l}{Average within one release}\\
		
		\midrule
		
		%Ant 1.3        & 0.416 & 0.342 & 0.392 & 0.316  & 0.335  & 0.363 & 0.202  & 1.079 \\
		%Ant 1.4        & 0.222 & 0.436 & NA    & 16.959 & 0.034  & 0.421 & 0.094  & 1.241 \\
		Ant 1.5          & \textbf{0.349} & \textbf{0.167} & 0.325 & \textbf{0.167}  & 0.307  & 0.176 & 0.127  & 0.597 \\
		%Ant 1.6        & 0.531 & 0.520 & 0.530 & 0.517  & 0.536  & 0.509 & 0.329  & 1.438 \\
		%Ant 1.7        & 0.458 & 0.491 & 0.463 & 0.495  & 0.456  & 0.500 & 0.369  & 0.881 \\
		%Camel 1.0      & NA    & 0.072 & 0.088 & 0.090  & -0.001 & 0.079 & 0.197  & 0.163 \\
		%Camel 1.2      & 0.385 & 0.908 & 0.287 & 1.036  & 0.380  & 0.931 & 0.440  & 1.236 \\
		%Camel 1.4      & 0.334 & 0.520 & 0.330 & 0.541  & 0.332  & 0.525 & 0.335  & 0.716 \\
		%Camel 1.6      & 0.302 & 0.781 & 0.242 & 0.773  & 0.267  & 0.754 & 0.370  & 1.246 \\
		Ivy 2.0          & \textbf{0.388} & 0.204 & 0.371 & \textbf{0.195}  & 0.374  & 0.196 & 0.160  & 0.401 \\
		%JEdit 3.2.1    & 0.614 & 1.328 & 0.532 & 1.776  & 0.551  & 1.314 & -0.337 & 3.120 \\
		%JEdit 4.0      & 0.532 & 0.697 & 0.531 & 0.722  & 0.508  & 0.727 & 0.478  & 0.974 \\
		JEdit 4.1        & 0.534 & \textbf{0.670} & \textbf{0.547} & 0.675  & 0.523  & 0.678 & 0.536  & 1.393 \\
		%JEdit 4.2      & 0.383 & 0.381 & 0.376 & 0.352  & 0.374  & 0.364 & 0.379  & 1.007 \\
		%Log4j 1.0      & 0.578 & 0.477 & 0.596 & 0.447  & 0.609  & 0.459 & 0.321  & 1.266 \\
		%Log4j 1.1      & 0.515 & 0.749 & 0.583 & 0.698  & 0.585  & 0.678 & 0.319  & 2.283 \\
		Log4j 1.2        & \textbf{0.426} & 0.924 & 0.359 & \textbf{0.879}  & 0.401  & 0.897 & 0.356  & 1.917 \\
		%Lucene 2.0     & 0.529 & 1.237 & 0.521 & 1.176  & 0.446  & 1.305 & 0.446  & 2.386 \\
		%Lucene 2.2     & 0.427 & 1.469 & 0.385 & 1.294  & 0.426  & 1.416 & 0.466  & 1.942 \\
		Lucene 2.4       & \textbf{0.564} & \textbf{1.384} & 0.554 & 1.425  & 0.550  & 1.446 & 0.550  & 2.328 \\
		%Poi 1.5        & 0.410 & 1.371 & 0.343 & 1.384  & 0.294  & 1.270 & 0.469  & 2.170 \\
		%Poi 2.0-RC1    & 0.243 & 0.212 & 0.253 & 0.194  & 0.217  & 0.205 & 0.249  & 0.349 \\
		Poi 2.5.1        & 0.615 & \textbf{0.840} & 0.557 & 0.880  & 0.593  & 0.857 & \textbf{0.687}  & 1.434 \\
		%Poi 3.0        & 0.567 & 0.843 & 0.429 & 0.824  & 0.550  & 0.862 & 0.539  & 1.276 \\
		%Synapse 1.0    & 0.325 & 0.210 & NA    & 0.837  & 0.286  & 0.215 & 0.154  & 0.321 \\
		%Synapse 1.1    & 0.528 & 0.454 & 0.412 & 0.526  & 0.514  & 0.461 & 0.495  & 0.728 \\
		%Synapse 1.2    & 0.472 & 0.583 & 0.376 & 0.720  & 0.471  & 0.582 & 0.424  & 0.917 \\
		%Velocity 1.4   & 0.573 & 0.562 & 0.587 & 0.607  & 0.323  & 0.637 & -0.021 & 4.283 \\
		Velocity 1.5     & 0.542 & 1.232 & \textbf{0.575} & \textbf{1.113}  & 0.503  & 1.197 & 0.508  & 7.554 \\
		%Velocity 1.6.1 & 0.465 & 0.989 & 0.397 & 0.950  & 0.332  & 0.973 & 0.386  & 4.216 \\
		%Xalan 2.4      & 0.341 & 0.328 & NA    & 0.341  & 0.361  & 0.332 & 0.301  & 0.492 \\
		%Xalan 2.5      & 0.384 & 0.656 & 0.293 & 0.651  & 0.316  & 0.657 & 0.339  & 0.764 \\
		%Xalan 2.6      & 0.533 & 0.666 & 0.511 & 0.645  & 0.528  & 0.656 & 0.561  & 0.650 \\
		Xalan 2.7        & 0.480 & 0.425 & 0.460 & 0.448  & 0.414  & 0.430 & \textbf{0.485}  & \textbf{0.400} \\
		%Xerces 1.2     & 0.469 & 0.367 & 0.448 & 0.359  & 0.332  & 0.386 & 0.365  & 0.550 \\
		%Xerces 1.3     & 0.420 & 0.622 & 0.332 & 0.629  & 0.171  & 0.609 & 0.202  & 0.899 \\
		Xerces 1.4.4     & \textbf{0.811} & 1.641 & 0.605 & \textbf{1.370}  & 0.576  & 1.568 & 0.746  & 1.793 \\
		%Xerces init    & 0.509 & 0.870 & 0.303 & 0.951  & 0.136  & 1.050 & 0.290  & 4.049 \\
		JDT              & 0.411 & 0.468 & \textbf{0.412} & \textbf{0.455}  & 0.403  & 0.471 & 0.403  & 0.727 \\
		EQU              & 0.544 & 0.717 & 0.554 & \textbf{0.648}  & \textbf{0.565}  & 0.651 & 0.489  & 1.454 \\
		%MYL            & 0.328 & 0.289 & 0.268 & 0.301  & 0.325  & 0.279 & 0.208  & 0.437 \\
		PDE              & \textbf{0.352} & 0.307 & 0.311 & 0.337  & 0.350  & \textbf{0.299} & 0.348  & 0.358 \\
		
		\bottomrule
	\end{tabular}
	\label{tab:textMethods}
\end{table}

Table~\ref{tab:textMethods} shows the average SCC and MAE across all evaluated releases in the top lines and some releases' best scores in the bottom lines. Best scores among different selection methods are marked in bold. In term of SCC, systems using Spearman method provided the best overall performance and stability. On average, they out performed the second best method, Pearson, by almost 7\%. On the other hand, MAE scores indicated that Spearman and MI selection methods have similar prediction errors. On average these two reduces error by almost 38\% comparing to the last method Pearson. In Table~\ref{tab:textMethods}, we could see that Spearman is not the best feature selection method for all evaluated releases but the differences are small. Paired t-tests confirmed that differences in performances between Spearman and other methods are statistically significant. However, the test could not confirm if, in term of MAE, Spearman is better than MI or PCC. It is important to note that using a feature selection method is better than not using: the results of no feature selection are often the worst.

In the second phase, we ran additional test using only Spearman as the selection method and extended the number of term features selected to include 3, 50, 100, and 200. This phase is designed to determine the optimal number of term features to be selected. We found that, selecting 5 term features seem provide the best performance. However the differences in predictive power between this setting and settings of 3, 10, and 20 are small and not statistically significant (confirmed using paired t-test). Our content-based features are project specific so different projects will have different optimal settings. It is important to note, however, that optimal settings for the number of selected term features are generally not more than 20.

\subsection{Experiment 2: Topic features}

This experiment was designed to confirm if topic modeling provide higher predictive power comparing to raw term features and which is the optimal number of extracted topics. We ran our system using topic features extracted by LDA, the numbers of topics were set to 5, 10, 15, 20, 30, 50, and 100. Table~\ref{tab:topicPackage} shows the predictive power of topic features and raw term features (without any feature selection method). In general topic features outperformed raw term features by a large margin of 28\% and 41\% in term of SCC and MAE scores respectively. This result suggests that topics extracted from source code are predictive of defects. %We will have a closer look at the detail topic extracted in Section~\nameref{sec:discussion} to confirm if they are representative of target systems' functionalities.

\begin{table}[t]
	\centering
	\caption{Average predictive power of topic and package features versus term and type features without feature selection}
	\begin{tabular}{l rr rr rr rr}
		\addlinespace
		\toprule
				
		Feature & \multicolumn{2}{c}{Topics} & \multicolumn{2}{c}{All terms} & \multicolumn{2}{c}{Packages} & \multicolumn{2}{c}{All types}\\
		Prediction measure & SCC & MAE & SCC & MAE & SCC & MAE & SCC & MAE\\

		\midrule

		%\multicolumn{9}{l}{Average across releases}\\

		%\midrule
		
		Across all releases             & \textbf{0.451} & \textbf{0.766} & 0.352  & 1.510 & \textbf{0.431} & \textbf{0.712} & 0.292 & 1.490 \\

		\midrule

		\multicolumn{9}{l}{Average within one release}\\
        
        \midrule

		%Ant 1.3        & 0.315 & 0.445 & 0.202  & 1.079 & 0.176 & 0.405 & 0.112 & 1.009 \\
		%Ant 1.4        & 0.170 & 0.417 & 0.094  & 1.241 & 0.442 & 0.333 & 0.245 & 0.771 \\
		Ant 1.5          & \textbf{0.358} & \textbf{0.219} & 0.127  & 0.597 & \textbf{0.294} & \textbf{0.213} & 0.212 & 0.371 \\
		%Ant 1.6        & 0.535 & 0.650 & 0.329  & 1.438 & 0.387 & 0.688 & 0.310 & 1.042 \\
		%Ant 1.7        & 0.473 & 0.618 & 0.369  & 0.881 & 0.412 & 0.602 & 0.274 & 1.010 \\
		%Camel 1.0      & 0.111 & 0.100 & 0.197  & 0.163 & 0.122 & 0.084 & 0.095 & 0.397 \\
		%Camel 1.2      & 0.384 & 1.029 & 0.440  & 1.236 & 0.539 & 0.965 & 0.374 & 1.972 \\
		%Camel 1.4      & 0.376 & 0.648 & 0.335  & 0.716 & 0.418 & 0.582 & 0.254 & 1.517 \\
		%Camel 1.6      & 0.371 & 0.868 & 0.370  & 1.246 & 0.436 & 0.808 & 0.197 & 2.501 \\
		Ivy 2.0          & \textbf{0.339} & \textbf{0.285} & 0.160  & 0.401 & \textbf{0.377} & \textbf{0.282} & 0.191 & 0.420 \\
		%JEdit 3.2.1    & 0.536 & 1.966 & -0.337 & 3.120 & 0.468 & 1.543 & 0.303 & 2.684 \\
		%JEdit 4.0      & 0.486 & 1.185 & 0.478  & 0.974 & 0.415 & 0.956 & 0.177 & 1.756 \\
		JEdit 4.1        & \textbf{0.549} & \textbf{0.963} & 0.536  & 1.393 & \textbf{0.454} & \textbf{0.828} & 0.249 & 1.525 \\
		%JEdit 4.2      & 0.398 & 0.549 & 0.379  & 1.007 & 0.354 & 0.451 & 0.134 & 1.047 \\
		%Log4j 1.0      & 0.512 & 0.665 & 0.321  & 1.266 & 0.407 & 0.666 & 0.201 & 2.039 \\
		%Log4j 1.1      & 0.565 & 0.896 & 0.319  & 2.283 & 0.406 & 0.889 & 0.302 & 2.186 \\
		Log4j 1.2        & \textbf{0.412} & \textbf{0.966} & 0.356  & 1.917 & \textbf{0.495} & \textbf{0.908} & 0.106 & 3.048 \\
		%Lucene 2.0     & 0.432 & 1.522 & 0.446  & 2.386 & 0.535 & 1.308 & 0.384 & 1.621 \\
		%Lucene 2.2     & 0.432 & 1.801 & 0.466  & 1.942 & 0.418 & 1.670 & 0.238 & 3.026 \\
		Lucene 2.4       & \textbf{0.559} & \textbf{1.549} & 0.550  & 2.328 & \textbf{0.458} & \textbf{1.657} & 0.333 & 2.797 \\
		%Poi 1.5        & 0.497 & 1.277 & 0.469  & 2.170 & 0.480 & 1.001 & 0.301 & 4.416 \\
		%Poi 2.0-RC1    & 0.278 & 0.221 & 0.249  & 0.349 & 0.294 & 0.203 & 0.122 & 0.622 \\
		Poi 2.5.1        & \textbf{0.724} & \textbf{0.677} & 0.687  & 1.434 & \textbf{0.766} & \textbf{0.614} & 0.607 & 1.373 \\
		%Poi 3.0        & 0.512 & 0.940 & 0.539  & 1.276 & 0.458 & 0.946 & 0.356 & 1.947 \\
		%Synapse 1.0    & 0.333 & 0.272 & 0.154  & 0.321 & 0.338 & 0.238 & 0.179 & 0.601 \\
		%Synapse 1.1    & 0.480 & 0.539 & 0.495  & 0.728 & 0.560 & 0.477 & 0.495 & 0.841 \\
		%Synapse 1.2    & 0.445 & 0.639 & 0.424  & 0.917 & 0.356 & 0.656 & 0.223 & 1.828 \\
		%Velocity 1.4   & 0.596 & 0.557 & -0.021 & 4.283 & 0.673 & 0.492 & 0.555 & 1.063 \\
		Velocity 1.5     & \textbf{0.625} & \textbf{1.051} & 0.508  & 7.554 & \textbf{0.590} & \textbf{1.069} & 0.396 & 1.729 \\
		%Velocity 1.6.1 & 0.437 & 0.992 & 0.386  & 4.216 & 0.340 & 1.003 & 0.307 & 1.955 \\
		%Xalan 2.4      & 0.382 & 0.354 & 0.301  & 0.492 & 0.300 & 0.344 & 0.206 & 0.513 \\
		%Xalan 2.5      & 0.410 & 0.598 & 0.339  & 0.764 & 0.422 & 0.593 & 0.317 & 0.889 \\
		%Xalan 2.6      & 0.581 & 0.565 & 0.561  & 0.650 & 0.476 & 0.614 & 0.549 & 0.792 \\
		Xalan 2.7        & \textbf{0.519} & \textbf{0.394} & 0.485  & 0.400 & 0.394 & \textbf{0.432} & \textbf{0.428} & 0.553 \\
		%Xerces 1.2     & 0.435 & 0.369 & 0.365  & 0.550 & 0.504 & 0.259 & 0.396 & 0.742 \\
		%Xerces 1.3     & 0.400 & 0.793 & 0.202  & 0.899 & 0.285 & 0.765 & 0.317 & 0.965 \\
		Xerces 1.4.4     & 0.667 & \textbf{1.699} & \textbf{0.746}  & 1.793 & \textbf{0.745} & \textbf{1.479} & 0.514 & 3.581 \\
		%Xerces init    & 0.558 & 0.884 & 0.290  & 4.049 & 0.618 & 0.905 & 0.266 & 1.746 \\
		JDT              & \textbf{0.441} & \textbf{0.521} & 0.403  & 0.727 & \textbf{0.416} & \textbf{0.501} & 0.327 & 0.771 \\
		EQU              & \textbf{0.621} & \textbf{0.714} & 0.489  & 1.454 & \textbf{0.483} & \textbf{0.776} & 0.438 & 1.071 \\
		%MYL            & 0.338 & 0.346 & 0.208  & 0.437 & 0.264 & 0.350 & 0.119 & 0.735 \\
		PDE              & \textbf{0.366} & 0.421 & 0.348  & \textbf{0.358} & \textbf{0.350} & \textbf{0.338} & 0.136 & 1.118 \\
		
		\bottomrule
	\end{tabular}
	\label{tab:topicPackage}
\end{table}

We also found from the experiment results that using 20 topics often provide the best overall predictive power. However the differences in term of performance between 20 topics and similar settings are small. The paired t-tests confirm that differences are not statistically significant. Topic modeling extracted latent topics that are corresponding to technical concerns (e.g. functionalities) of specific target projects, it is ideal that the number of topics matches the number of functionalities. However, each of target projects would have different number of functionalities, which mean the optimal number of topics will change from project to project. The result suggests that there is no clear global optimal setting for the number of extracted topics and this setting is generally best set around 20 (performance of using 100 topics are significantly lower).

\subsection{Experiment 3: Type features and type feature selections}

Experiment 3 was conducted using the same process as Experiment 1, it was used to investigate the effectiveness of type features in defect prediction task by measuring their predictive powers and errors. In the first phase, comparing different feature selection methods, we ran the system using three methods of selection (Spearman, Pearson and MI) with the number of type features selected of 5, 10, and 20. %We also ran experiment in both count and binary mode.

%\begin{figure*}[htb]
%	\caption{Type feature in count and binary mode}
%	\centering
%	\subfloat[Spearman]{\includegraphics[width = 0.3\textwidth]{Results/Semantic_bin_vs_count_0-5-10-20_LRS}}
%	\subfloat[MAE]{\includegraphics[width =0.3\textwidth]{Results/Semantic_bin_vs_count_0-5-10-20_LRMAE}}
%	\subfloat[Relative Cost Effectiveness]{\includegraphics[width =0.3\textwidth]{Results/Semantic_bin_vs_count_0-5-10-20_LRE2.pdf}}
%	\label{semaBinVSCount}
%\end{figure*}

%The result in figure\ref{semaBinVSCount} shows that binary mode work very well with type feature. We can see that in term of Spearman correlation measure, binary mode work significantly better than count mode \todo{by how much, give \%}. In two other performance measure the two are relatively similar. In this case we will chose to use binary mode as it is simpler and performs better in term of the most relevant measure, Spearman correlation.

\begin{table}
	\centering
	\caption{Average predictive power of different type features selection methods}
	\begin{tabular}{l rr rr rr rr}
		\addlinespace
	 
	 	\toprule
				
		Selection method & \multicolumn{2}{c}{Spearman} & \multicolumn{2}{c}{Pearson} & \multicolumn{2}{c}{MI} & \multicolumn{2}{c}{All features}\\
		Measure & SCC & MAE & SCC & MAE & SCC & MAE & SCC & MAE\\

		\midrule

		%\multicolumn{9}{l}{Average across releases}\\

		%\midrule
		
		Across all releases         & \textbf{0.443} & 0.688 & 0.417 & 0.913 & 0.425 & \textbf{0.686} & 0.292 & 1.490 \\

		\midrule

		\multicolumn{9}{l}{Average within one release}\\
        
        \midrule

		%Ant 1.3        & 0.334          & 0.344          & NA             & 0.342 & 0.290          & 0.356          & 0.112 & 1.009 \\
		%Ant 1.4        & 0.406          & 0.329          & 0.348          & 0.357 & 0.390          & 0.335          & 0.245 & 0.771 \\
		Ant 1.5          & 0.347          & \textbf{0.162} & \textbf{0.392} & 0.165 & 0.312          & 0.173          & 0.212 & 0.371 \\
		%Ant 1.6        & 0.446          & 0.630          & 0.372          & 0.715 & 0.445          & 0.629          & 0.310 & 1.042 \\
		%Ant 1.7        & 0.424          & 0.533          & 0.412          & 0.537 & 0.412          & 0.541          & 0.274 & 1.010 \\
		%Camel 1.0      & NA             & 0.228          & NA             & 0.210 & 0.161          & 0.074          & 0.095 & 0.397 \\
		%Camel 1.2      & 0.358          & 0.982          & NA             & 1.749 & 0.360          & 0.972          & 0.374 & 1.972 \\
		%Camel 1.4      & 0.333          & 0.531          & NA             & 1.014 & 0.235          & 0.559          & 0.254 & 1.517 \\
		%Camel 1.6      & 0.298          & 0.723          & NA             & 4.067 & 0.290          & 0.742          & 0.197 & 2.501 \\
		Ivy 2.0          & 0.232          & 0.254          & 0.234          & 0.254 & \textbf{0.255} & \textbf{0.249} & 0.191 & 0.420 \\
		%JEdit 3.2.1    & 0.577          & 1.607          & NA             & 1.862 & 0.608          & 1.626          & 0.303 & 2.684 \\
		%JEdit 4.0      & 0.446          & 0.989          & NA             & 1.462 & 0.451          & 1.005          & 0.177 & 1.756 \\
		JEdit 4.1        & \textbf{0.552} & \textbf{0.779} & 0.326          & 1.143 & 0.526          & 0.793          & 0.249 & 1.525 \\
		%JEdit 4.2      & 0.395          & 0.458          & NA             & 0.593 & 0.420          & 0.467          & 0.134 & 1.047 \\
		%Log4j 1.0      & 0.477          & 0.587          & 0.416          & 0.674 & 0.450          & 0.586          & 0.201 & 2.039 \\
		%Log4j 1.1      & 0.632          & 0.746          & 0.608          & 0.821 & 0.634          & 0.732          & 0.302 & 2.186 \\
		Log4j 1.2        & \textbf{0.411} & \textbf{0.932} & 0.270          & 0.968 & 0.397          & 0.936          & 0.106 & 3.048 \\
		%Lucene 2.0     & 0.572          & 1.154          & NA             & 1.248 & 0.535          & 1.240          & 0.384 & 1.621 \\
		%Lucene 2.2     & 0.378          & 1.532          & NA             & 1.678 & 0.387          & 1.554          & 0.238 & 3.026 \\
		Lucene 2.4       & 0.453          & 1.555          & 0.295          & 1.861 & \textbf{0.486} & \textbf{1.516} & 0.333 & 2.797 \\
		%Poi 1.5        & NA             & 1.188          & 0.248          & 1.317 & 0.450          & 1.087          & 0.301 & 4.416 \\
		%Poi 2.0-RC1    & 0.261          & 0.280          & NA             & 0.308 & 0.261          & 0.198          & 0.122 & 0.622 \\
		Poi 2.5.1        & \textbf{0.700} & 0.690          & 0.674          & 0.700 & 0.693          & \textbf{0.637} & 0.607 & 1.373 \\
		%Poi 3.0        & 0.424          & 0.979          & 0.138          & 0.981 & 0.376          & 0.985          & 0.356 & 1.947 \\
		%Synapse 1.0    & 0.245          & 0.204          & NA             & 0.262 & 0.316          & 0.202          & 0.179 & 0.601 \\
		%Synapse 1.1    & 0.539          & 0.439          & NA             & 0.605 & 0.544          & 0.442          & 0.495 & 0.841 \\
		%Synapse 1.2    & 0.419          & 0.618          & NA             & 0.819 & 0.414          & 0.623          & 0.223 & 1.828 \\
		%Velocity 1.4   & 0.540          & 0.546          & 0.536          & 0.534 & 0.557          & 0.530          & 0.555 & 1.063 \\
		Velocity 1.5     & 0.547          & 1.111          & 0.554          & \textbf{1.085} & \textbf{0.555}          & 1.096          & 0.396 & 1.729 \\
		%Velocity 1.6.1 & 0.478          & 0.847          & 0.478          & 0.906 & 0.474          & 0.853          & 0.307 & 1.955 \\
		%Xalan 2.4      & 0.294          & 0.323          & 0.189          & 0.383 & 0.290          & 0.324          & 0.206 & 0.513 \\
		%Xalan 2.5      & 0.327          & 0.635          & 0.170          & 1.144 & 0.316          & 0.639          & 0.317 & 0.889 \\
		%Xalan 2.6      & 0.503          & 0.608          & 0.489          & 0.608 & 0.508          & 0.604          & 0.549 & 0.792 \\
		Xalan 2.7        & 0.515          & 0.346          & 0.415          & 0.411 & \textbf{0.518} & \textbf{0.345} & 0.428 & 0.553 \\
		%Xerces 1.2     & 0.514          & 0.285          & 0.498          & 0.295 & 0.512          & 0.290          & 0.396 & 0.742 \\
		%Xerces 1.3     & 0.385          & 0.547          & NA             & 0.812 & 0.400          & 0.562          & 0.317 & 0.965 \\
		Xerces 1.4.4     & \textbf{0.677} & \textbf{1.397} & 0.323          & 1.752 & 0.487          & 1.535          & 0.514 & 3.581 \\
		%Xerces init    & 0.290          & 1.094          & NA             & 1.190 & 0.215          & 1.089          & 0.266 & 1.746 \\
		JDT              & \textbf{0.458} & \textbf{0.409} & 0.404          & 0.437 & 0.453          & 0.410          & 0.327 & 0.771 \\
		EQU              & 0.522          & \textbf{0.649} & 0.499          & 0.726 & \textbf{0.537} & 0.651          & 0.438 & 1.071 \\
		%MYL            & 0.209          & 0.310          & NA             & 0.356 & 0.221          & 0.308          & 0.119 & 0.735 \\
		PDE              & \textbf{0.301} & \textbf{0.338} & 0.211          & 0.779 & 0.285          & \textbf{0.338} & 0.136 & 1.118 \\
		
		\bottomrule
	\end{tabular}
	\label{tab:semanticMethods}
\end{table}

Overall, Spearman has the best predictive power (SCC) among  selection methods for type features. Spearman on average perform better than MI, the second best method, by over 4\% in term of SCC. MI on average has lower prediction error comparing to Spearman but only with the reduction of less than half of a percent. In the list of sampled projects, we can see that Spearman is the better selection method on most cases.

We conducted the second phase of this experiment to investigate the relation between number of selected type features with the predictive power of our system. To do this we used the established method of choice Spearman with different number of selected features of 50, and 100 (results for setting of  5, 10, and 20 were carried over from previous phase). We found that selecting 10 type features often provides the best overall predictive power, however the differences between settings are small.

\subsection{Experiment 4: Package features}

Experiment 4 was perform to evaluate our last feature type, package features. Feature vectors were extracted based on functional (non-empty) packages. Since these functional packages are organizations of member classes, they would provide abstract concepts of those classes. We assume that in a well designed software, these functional package would contain classes of different functionalities. PCA was applied to remove redundant information and reduce noise.

We compared package features with raw type features, without any feature selection technique, to prove that extra levels of abstraction improve defect prediction performance. As Table~\ref{tab:topicPackage} shows, the overall predictive power was improved significantly, by almost 48\%, and the average error dropped by over 52\%. This result suggests that package organization contains significant amount of information in term of software modularity which provide our system a way to categorize type features and that help improve defect prediction performance.

\subsection{Experiment 5: The best prediction system}

To find the best prediction system we decided to combine all features using their best overall settings with the assumption is that each type of feature will contain a different aspect of the target system's semantic complexity. PCA was applied on input vectors to reduce noise and overlapping data, the variance thread-hold was set to 90\%.

\begin{table}[t]
	\centering
	\caption{Average predictive power of different feature types}
	\begin{tabular}{@{}l r r r r r r r @{}}
		\addlinespace
		\toprule
		Feature type	 & Combined & Term & Topic & Type & Package & CKOO & LOC\\
		
		\midrule
		
		Mean SCC & \textbf{0.4624} & 0.4621 & 0.4550 & 0.4426 & 0.4315 & 0.3362 & 0.3357 \\
		Mean MAE   & \textbf{0.6733} & 0.6806 & 0.7560 & 0.6880 & 0.7117 & 0.7638 & 0.7686 \\

		\bottomrule
	\end{tabular}
	\label{tab:overall}
\end{table}

To maintain consistency, we ran our combination test using best overall settings as well as neighboring setting values. We ran experiments using the Spearman feature selection on both term and type features with number of feature selected set to 3, 5, 10 and 5,10, 20 respectively, number of extracted topics was set to 5, 10, 15, 20, and 30. Because we only want to confirm if combining different types of feature does create a better predictive system, performance of combined features is compared directly to individual feature types using the same settings.

Table~\ref{tab:overall} shows the predictive power of systems that use combined features, our individual feature types, and two baseline systems using traditional metrics, CK and LOC. The result suggests that all four of our proposed feature types outperform the baseline systems in both
prediction measures SCC and MAE. Among newly proposed feature types, term features create the biggest predictive power improvement of 38\% and the largest prediction error reduction of 11\% over the best conventional metrics (CKOO). 
%In Figure~\ref{fig:overall}, we could also see that our proposed features have more stable performance than base metrics, this is shown by the narrower box plots' boundaries of our systems.
Table~\ref{tab:overall} shows that combining all of our features does create a better prediction system, but the improvements are insignificant. On average, combined features improve SCC by only 0.4\% and reduce MAE by only 1\%. Using such an simple combination method might be the reason why our combined features did not show larger improvement. More sophisticated methods might create better results, we leave this to our future work.

\subsection{Case study}

The evaluation results suggest that the content-based features extracted from source code provide substantial improvements in defect prediction. We investigated a representative subject system JEdit 3.2.1 as a case study into those features.

\subsubsection{Topic features}

To verify in details if topics extracted from source code using topic modeling are representative of the functionality of the target system, we extracted the most relevant source files for each topic, i.e. ones contain most words assigned to that topic. That list could help us infer the corresponding functionality of each topic. %Table~\ref{tab:topicFeature} shows the five extracted topics for JEdit 3.2.1 along with performance measurements to evaluate the effectiveness of each topic. We included the labels for those topics, inferred by our best effort.

We found five topics corresponding to five functionalities of JEdit: 1. \textit{Text parsing}, 2. \textit{Interpreter}, 3. \textit{View and edit components}, 4. \textit{Data model}, and 5. \textit{Graphical User Interface (GUI) components}. JEdit is a code editor for programmers, i.e. its main functions provide a developing environment. As an editor, most operations will involve user interaction to view, edit, copy, and delete text files. Components that are responsible to provide such functions are likely to be used extensively and will be subjected to multiple changes that are likely to introduce defects. We found that Topic 3 involving view and edit functionality has high predictive power to defects. For example, it has a Spearman ranked correlation coefficient of 0.701 (higher than all selected term and type features). Topic 5 involving GUI components is also highly predictive of defects with a ranked correlation of 0.610 (higher than all selected type features). In contrast, Topic 2 and Topic 5 has nearly zero correlation (-0.135 and 0.157, respectively).

\subsubsection{Package features}

Our package features are extracted based on an assumption that a well-designed software system will have a generally good level of modularity and the package structure of this system will reflect its modular organization. 

We found that two major packages of JEdit 3.2.1, \code{org.gjt.sp.jedit} and \code{org.gjt.sp.util} have high correlation to defect (and higher than all listed baseline metrics). They are also appear exclusive in most defect files. For example, 90\% of top defective files refer to \code{org.gjt.sp.util}, while it is referred only 11\% in files having no defects. Their sub-packages, like \code{org.gjt.sp.jedit.browser} or \code{org.gjt.sp.jedit.syntax} appear even more exclusive in top defective files. This explain why package features are predictive of defects. 

%top 10 features are generally predictive of defect, with relatively larger R-squared scores. Feature browser, search, and syntax have relatively high defect density estimation (LR Coef) with significant P-value, this means that usages of their types are generally defect-prone. These three packages provide JEdit with classes to browse file system, to search and replace, and to perform syntax highlighting, all of which are core functionalities of JEdit. 

It is interesting to observe that package \code{bsh} (BeanShell) has a negative correlation with defect counts, i.e.  source files use classes belonging to this package are more likely to have of no defects. We found that that package provides scripting capabilities for Java. However, due to the inclusion of JavaScript in JDK, its development has been discontinued since 2005. Although many components of JEdit still use this package, due to this discontinued development, they are likely legacy code, i.e. having no changes, and thus are likely to have no newly injected defects.

We further studied package features in Ant 1.4, a tool for compiling and building executable code for software projects. We found that, three packages \code{org.apache.tools.ant.taskdefs}, \code{org.apache.tools.ant.taskdefs.rmic}, and \code{org.apache.tools.ant.taskdefs.condition} have the highest correlation to defect among all package features. Their correlation is higher than that of all baseline metrics. That is reasonable because they contain core functionality of Ant, a build tool. For example, package \code{taskdefs} is for defining build tasks and \code{taskdefs.condition} is for specifying execution conditions on those build tasks. In contrast, \code{org.apache.tools.zip} and \code{org.apache.tools.tar} are external functionality for processing compressed files, and they have nearly zero correlation to defects. It is expected that components providing core functionality are more defect-prone than others because they often have more active development activities and stricter requirements.

\subsubsection{Defect distribution}
We observed that JEdit 3.2.1 has very skew defect distribution where a small number of files have significantly large numbers of defects, while the rest contain few to no defects. For example, file \code{JEditTextArea.java} contains 45 post-release defects, accounted for 12\% of the total defects and the top 10 most defective files account for 47\% of total defects. On the other hand, other systems like Ant 1.4 have  more balanced defect distributions. The most defective file in Ant 1.4 has only 3 defects and the top 10 most defective files account for only 36\% of total defects. 

Systems with skew defect distributions have higher prediction performance than the more balanced ones. For example, the best SCC of JEdit 3.2.1 is 0.635 (over experiment settings for feature types and selection methods) while that of Ant 1.4 is lower at 0.441. The reason is possibly due to the training process of Linear Regression (LR) models. LR models are trained to minimize the sum of squared residuals. Because outliers, like ones with unusual high defect counts, would have high impact on that sum, the model is likely to be trained to reduce the errors when predicting those files. In other words, it will predict top defective files more accurately, and thus, improves the ranked correlation (which is also strongly influenced by files with high defect counts). If the defect distribution is more balanced, then the prediction model will less focus on files with high defect counts, thus, predicts them with larger errors which leads to a lower SCC score.
\section{Related work}

Defect prediction has attracted great research interest in software engineering. Researchers have searched for: 1) software metrics that are predictive of defect-proneness, and 2) prediction models that deliver accurate results.

%\subsection{Software metrics}

Software metrics used in defect prediction studies mainly fall into three categories: product metrics (e.g. static code metrics), process metrics (e.g. code churn and previous bugs), and socio-technical metrics (e.g. developers).

One of the earliest code metrics used in software defect prediction is lines of code (LOC). Simple, easy to extract, frequently used, this metrics is still being discussed to this day. In a number of studies, LOC has been reported to correlate with the number of faults and performs quite well, although other studies show that LOC has only modest predictive power. In general, LOC appears to be useful in predicting software defects (a survey of LOC's use can be found in \cite{hall2011}). Beside LOC, popular complexity metrics such as McCabe's cyclomatic complexity have been shown to be useful in predicting software defects. \cite{khoshgoftaar03} used 16 code complexity metrics to predict defects in a large legacy system for telecommunications and achieved accuracy of almost 80\% when using modules from one release as training data to predict fault in the consecutive release of the system. In addition to general code metrics, object-oriented design measures have been widely used in defect prediction studies. \cite{basili96} investigated the usefulness of of the Chidamber and Kemerer (CK) metrics in predicting bugs. They found five metrics correlating with the defect count: weighted methods per class, coupling between objects, the depth of inheritance, the number of children, and the response for a class. In a later study \cite{briand99} showed that coupling between objects, lack of cohesion among methods, and response for a class were highly predictive of fault-proneness of a class. Overall, OO metrics have been reported to outperform general complexity metrics~\cite{hall2011}.

Process metrics such as code churn, the number of changes, previous bugs, are extracted from the software development history. \cite{graves00} studied the change history and found that large and recent changes contributed the most to defects. \cite{nagappan05} showed that system defect density could be predicted using a set of relative code churn measures that relate the amount of churns to other variables such as component size and the temporal extent of a churn. They later found that change bursts (i.e. frequently changed code) could be used has good predictors of bugs. \cite{hassan-icse09} proposed to use the entropies of changes as measures of code change complexity and found that such measures outperformed the absolute numbers of changes.

There are arguments in favor of both product and process metrics. While process metrics appear to deliver more accurate predictions than product metrics, the latter is easier to obtain because it is derived from the code itself and does not require any information from the development process. When both types of metrics are available, it is useful to combine them for better predictions. In addition to these two categories of metrics, socio-technical metrics that are computed from the information on organization structure, developers, and social networks have also been shown to be useful in predicting effects.

%\subsection{Content-based features for defect prediction}

Some content-based features has been explored previously for defect prediction. \cite{tan.ase13} use all words and special operators as features in their classifier of buggy changes. However, as they did not use any feature selection/reduction techniques and nor provide any detailed analysis on the defect-proneness and predictive power of their features, our work is a complementary treatment and contains a deeper analysis of those \emph{textual features}.

\cite{chen} reports an empirical study using topics extracted from source code to \emph{explain} defects. As their study focuses on explaining defect-proneness using topics, they did not perform defect prediction nor analyze the effect of feature selection on the performance of the prediction tasks. Thus, our work provides a deeper study of the relationship between topics and defects. \cite{nier11} only study topic features on one subject system and one configuration of topic modeling. In this paper, we explored more types of features and ran experiments on many more configurations and subject systems.

\section{Conclusions}
In this paper, we explore a content-based approach for defect prediction. Our approach extracts text, topic, type, and package features from the textual and semantic content of the source code and uses them as defect predictors. Empirical evaluation shows that i) our content-based features are predictive of defect-proneness and have higher predictive power than traditional code metrics. In addition, selecting, reducing, and combining those features provides better prediction performance than using them individually.

\emph{Future work.} In this paper we design the features based on direct processing of textual and semantic content of source code. However this method of feature engineering might not cover all possible aspects of source code semantics. Recent advanced unsupervised text processing and pre-trained language models that produce distributed representation of terms and sentences was successfully applied to the natural languages. Thus, it is a natural next step to apply these techniques directly to source code. In addition, we could extend that concept to learn the distributed representations of more abstract code features such as data types, functions, classes or entire programs. These new representations could be the key to improve the performance of defect prediction systems.

\bibliography{defect}
\end{document}